\input harvmac
\def \h {\hat}
\def\ff{ {\rm flat} }

\def \s {\sigma}
\def \p {\phi}
\def \h {\hat }
\def \ha {\half}
\def \ov {\over}

\def \four{{\textstyle {1\ov 4}}}
\def \a {\alpha}
\def \lr { \lref}
\def\ep{\epsilon}

\def\vp {\varphi}

\def\m{\mu}\def\n {\nu}\def \G {\Gamma}
\def \del {\partial}

\def \bt {{\bar \theta}}
\def \ha{{\textstyle{1\over 2}}}

\def \DD {{\cal D}}
\def \a {\alpha}
\def \b {\beta}
\def \zeta {\zeta}
\def \s {\sigma}
\def \p {\phi}
\def \m {\mu}
\def \n {\nu}
\def \vp {\varphi }

\def \t {\theta}
\def \td {\tilde }

\def \sm {$\s$-model\ }

\def \inv {^{-1}}
\def \ov {\over }
\def \four{{\textstyle{1\over 4}}}

\def \third {{\textstyle{1\over 3}}}
\def\g {\Gamma}

\def   \td {\tilde }
\def\k{\kappa}

\def\np {  Nucl. Phys. }
\def \pl { Phys. Lett. }
\def \mpl { Mod. Phys. Lett. }
\def \prl { Phys. Rev. Lett. }
\def \pr  { Phys. Rev. }
\def \cqg { Class. Quantum Grav.}

\baselineskip8pt
\Title{\vbox
{\baselineskip 6pt
{\hbox {Imperial/TP/97-98/038  }}{\hbox{ NSF-ITP-98-048 }}
{\hbox{hep-th/9804076}} 
{\hbox{   }}
}}
{\vbox{
\centerline {
Green-Schwarz superstring  action in a curved }
 \centerline {    }
 \centerline {magnetic 
 Ramond-Ramond    background  } 
\vskip4pt }}

\centerline  {J.G. Russo$^1$\footnote {$^*$} 
{e-mail address: jrusso@ic.ac.uk} 
and 
A.A. Tseytlin$^{1,2}$\footnote{$^{\star}$}
{\baselineskip8pt e-mail address: tseytlin@ic.ac.uk}\footnote{$^{\dagger}$}{\baselineskip8pt
Also at  Lebedev  Physics
Institute, Moscow.} 
}



\medskip

\medskip
\smallskip\smallskip
\centerline {$^1$\it  Theoretical Physics Group, Blackett Laboratory,}
\smallskip
\centerline {\it  Imperial College,  London SW7 2BZ, U.K.}
\medskip
\centerline {$^2$\it Institute of   Theoretical Physics, University of California,}
\smallskip
\centerline {\it  Santa Barbara, CA93106, USA}
\bigskip\bigskip
\centerline {\bf Abstract}
\medskip
\baselineskip10pt
\noindent
\medskip
We derive the complete covariant action for the
 type IIA  superstring in a simple $D=10$ background which 
 represents a 7-brane with a magnetic  Ramond-Ramond   vector field
 (and  is U-dual to
  the Kaluza-Klein Melvin solution).  
This curved background can be obtained by dimensional reduction from a flat
(but topologically non-trivial) $D=11$ space-time.
The  action of a supermembrane
propagating in this flat 
 $D=11$ space  is straightforward to write down.
The explicit form of the 
 superstring action is then obtained 
by double dimensional reduction 
of  the  supermembrane action. In the light-cone gauge
the  action contains only quadratic and quartic terms in fermions.

\Date {April 1998}

\noblackbox \baselineskip 16pt plus 2pt minus 2pt 

\def \k {\kappa}

\lr \chap {A.H. Chamseddine, \np B185 (1981) 403;
E. Bergshoeff, M. de Roo, B. de Wit and P. van Nieuwenhuizen, \np B195 (1982) 97; G.F. Chaplin and N.S. Manton, \pl B120 (1983) 105.} 

\lr \witten {E. Witten, \np B443 (1995) 85, hep-th/9503124. }

\lr \ft { E.S. Fradkin  and A.A. Tseytlin, \pl B160 (1985) 69.  }

\lr \ftt { E.S. Fradkin  and A.A. Tseytlin, \pl B158 (1985) 316;  
 A.A. Tseytlin,  \pl B208 (1988) 221. }

\lr\dewi{
 B. de Wit, J. Hoppe and  H. Nicolai, 
 Nucl. Phys.B305 (1988) 545.}

\lr \towns{
P.K. Townsend,
{``Eleven-dimensional supermembrane revisited"},
 Phys. Lett. { B350} (1995) 184,
hep-th/9501068.}

\lr \gibma { G.W.  Gibbons and  K. Maeda, \np B298 (1988) 741.}

\lr\gib{G.W.  Gibbons, in: {``Fields and Geometry"}, Proceedings of the 22nd
Karpacz Winter School of Theoretical Physics, ed. A. Jadczyk (World Scientific, Singapore,  1986).}

\lr \gaun {F. Dowker, J.P. Gauntlett, G.W. Gibbons and G.T. Horowitz, ``The
decay of magnetic fields in Kaluza-Klein theory",
Phys. Rev.D52  (1995) 6929-6940,  hep-th/9507143.}

\lr\dowo{  F. Dowker, J.P. Gauntlett, D.A. Kastor and J. Traschen,
\pr D49 (1994) 2909; F. Dowker, J.P. Gauntlett, S.B. Giddings and G.T. Horowitz, \pr D50 (1994) 2662.}

\lr \gs{ M.B. Green  and J.H. Schwarz, \pl B149 (1984) 117;
\pl B151 (1985) 21; \np B255 (1985) 93. }

\lr \callan {C.G. Callan, C. Lovelace, C.R. Nappi and S.A. Yost, \pl B206 (1988) 41; \np B308 (1988) 221.}

\lr \green {M.B. Green, J.H.  Schwarz and E.  Witten, {\it Superstring Theory} (Cambridge U.P., 1988).}

\lr \green {M.B. Green, J.H.  Schwarz and E.  Witten, {\it Superstring Theory}
(Cambridge U.P., 1988).}
\lr \nils{ B.E.W. Nilsson and A.K. Tollsten, \pl 169 (1986) 369; R. Kallosh, 
Phys. Scr. T15 (1987) 118.  }
\lr \tsss{ } 
\lr \alw{L. \'Alvarez-Gaum\'e and E. Witten, \np B234 (1983) 269.     }

\lr \crem { E. Cremmer  and S. Ferrara, \pl B91 (1980) 61. }

\lr \bri { L. Brink  and P. Howe, \pl B91 (1980) 384. }

\lr\rutse{J.G.~Russo and A.A.~Tseytlin, \np B449 (1995) 91, hep-th/9502038.} 

\lr \duf { M.J. Duff, P.S. Howe, T. Inami and K.S. Stelle, 
\pl B191 (1987) 70. }

\lr\bdewit {B. de Wit, K. Peeters and J. Plefka, 
 ``Superspace geometry for supermembrane backgrounds", 
hep-th/9803209.}

\lr\poly{D. Polyakov,  ``RR - dilaton interaction in type II superstring", RU-95-85,  hep-th/9512028.}

\lr \hul{E. Bergshoeff, C.  Hull  and T. Ortin, \np  B451 (1995) 547, hep-th/9504081.}

\lr\gups{Gupser}

\lr \seza {M. Huq and M.A. Namazie, \cqg 2 (1985) 293, 597 (E); 
F. Giani and M. Pernici, \pr D30 (1984) 325; I.C. Campbell and P.C. West, \np B243 (1984) 112; S.J. Gates, Jr., J. Carr and R. Oerter, \pl  B189 (1987) 68.}

\lr \sezb{ J.H. Schwarz, \np B226 (1983) 269; P.S. Howe and P.C. West, \np B238 (1984) 181.}

\lr \gs{ M.B. Green and J.H. Schwarz, \pl B136 (1984) 376; \np B243 (1984) 285.}

\lr \wiit{E. Witten, \np B266 (1986) 245;  
J.J. Atick, A. Dhar and B. Ratra, \pl B169 (1986) 54; 
R. Kallosh, Phys. Scr. T15 (1987) 118;
M.T. Grisaru and D. Zanon, }

\lr \mizi{M.T. Grisaru, P.S. Howe, L. Mezincescu, B.E.W. Nilsson and P.K. Townsend, \pl B162 (1985) 116.  }
\lr \achu {A. Ach\'ucarro, P. Kapusta and K.S. Stelle, \pl B232 (1989) 302.}

\lr \berg{ E. Bergshoeff, E. Sezgin and P.K. Townsend, \pl B189 
(1987) 75; Ann. of Phys.  185 (1988) 330. }

\lr \pkt {P.K. Townsend, \pl B350 (1995) 184.}
\lr \cree{E. Cremmer, B. Julia and J. Scherk, \pl B76 (1978) 409.}

\lr \ber {N. Berkovits and W. Siegel,
 ``Superspace effective actions for 4D compactifications of heterotic
  and type II superstrings", Nucl. Phys. B462 (1996)  
  213, hep-th/9510106;
  N.~Berkovits, ``Extra dimensions in superstring theory", 
  Nucl. Phys. B507 (1997) 731, hep-th/9704109.
  }

\lr\tsey{A.A.~Tseytlin, 
``On dilaton dependence of type IIB superstring  action",
\cqg 13  (1996)  L81, hep-th/9601109.}

\lr\melvin{J.G. Russo and A.A. Tseytlin,
 ``Magnetic flux tube models in superstring theory", 
 \np B461 (1996) 131, hep-th/9508068;
 A.A. Tseytlin, 
 ``Closed superstrings in magnetic field:
 instabilities and supersymmetry breaking", 
 Nucl. Phys. Proc. Suppl. 49 (1996) 338, 
hep-th/9510041.  }

\lr \wit{S.J. Gates,  Jr.,  and B. Zwiebach, \np B238 (1984) 99;
 E. Witten, \pl B155 (1985) 151;
R.E. Kallosh, \pl B159 (1985) 111.}
\lr \schw {J.H. Schwarz, Phys. Rep. 89 (1982) 223.}
 \lr \gatt{S. Bellucci, S.J. Gates, Jr., B. Radak, P. Majumdar
   and Sh. Vashakidze,  \mpl  A4 (1989) 1985;
S.J. Gates, Jr., P. Majumdar, B. Radak and  Sh. Vashakidze,
    \pl   B226 (1989) 237;  B. Radak and Sh. Vashakidze, \pl B255 (1991) 528.}

\lr\pol{ J. Polchinski, \prl  75 (1995) 4724.}

\lr\mald{N. Itzhaki, J. Maldacena, J. Sonnenschein and
  S. Yankielowicz, {``Supergravity and the large N
limit of theories with sixteen supercharges"},
 hep-th/9802042;
  N. Itzhaki, A.A. Tseytlin   and S. Yankielowicz,
``Supergravity solutions  for  branes
localized within branes", hep-th/9803103.}



1. One interesting    problem in superstring theory 
is to understand how
 fundamental strings propagate
  in the presence of background  fields corresponding 
  to the Ramond-Ramond (RR) sector
of type II superstring spectrum.
This is important, in particular, in view of 
 the  presence  of  D-branes  \pol\ in the theory. 
It is natural to try to address    this question 
 using  the  Green-Schwarz (GS) formulation  \gs\ in which the
 string action  in a  RR background is local
and supersymmetric. 
The  leading-order terms in the coupling of GS  superstring to RR fields 
can be constructed \ft\  in the light-cone gauge
using 
the  known  light-cone gauge  GS vertex operators \schw.
The formal  superspace 
expressions for  the $\k$-invariant  type II  
GS superstring actions in   generic on-shell
 $N=2,  D=10$ supergravity backgrounds were found   
  in \refs{\mizi, \duf} (and studied in   
  \gatt),  but the explicit   component form of the actions
    was not  worked out. 
A  different   approach to the construction 
of manifestly $N=2, D=4$ supersymmetric world sheet \sm   for   compactified $D=4$
Type II  superstring    was    presented in \ber.   
The dilaton dependence  of RR coupling terms in GS action was 
explicitly demonstrated 
in \tsey. 
Certain leading component 
terms in $D=11$ membrane action \berg\ 
and thus \duf\ in $D=10$ string action
were recently determined  
in  \bdewit.

 Given that finding  
 the complete expression for the 
 covariant superstring action  in terms of
  the component 
fields $(x,\t)$ to all orders in
$\t$'s  in  a generic  supergravity background  is 
obviously a complicated  problem,\foot{GS action is complicated even in curved
NS-NS backgrounds, 
but in the absence of RR fields  one  may use the 
well-understood 
NS sigma-model
representation for the superstring action.
}
 one may  first 
try to solve it  for  some special 
cases  of the  RR backgrounds. 
In this paper we shall 
consider a simple example
of the  RR background that solves the type IIA supergravity 
equations of motion. It  
represents a  (non-supersymmetric)
magnetic 7-brane of type IIA theory  which is
the RR  analogue  (actually,  U-dual) of the NS-NS 
 Kaluza-Klein Melvin (`flux tube') 
background. 
The latter was previously discussed at the 
field theory level
   \refs{\gib \gibma \dowo -\gaun } 
and   at the 
string theory level \refs{\melvin }\ 
(the corresponding 
   type II superstring  model   is exactly soluble 
   so that its mass spectrum and
partition function  can be explicitly determined \melvin).

This background can be obtained 
by  dimensional reduction from  a  simple eleven-dimensional space-time
which is  {\it flat}
 (and  thus should 
be an exact solution of M-theory)
but topologically non-trivial. 
 Its non-trivial 3-dimensional part  is obtained 
by  factorizing $(R^2)_{r,\vp} \times (S^1)_y$ over the group generated by
 translations in  two angular directions.\foot{
In the coordinates where $ds^2=
dr^2 + r^2 d\t^2 + dy^2 $  $(\t=\vp + q y$) one 
identifies  the points $(r,\t, y) = (r, \t + 2\pi n   + 2\pi qRm, y+
2\pi R m)$ \ ($n,m=$integers), i.e. combines the shift  by $2\pi R$ in $y$ with
a rotation by an arbitrary angle $2\pi qR$ in the 2-plane.
The fixed $r$ section is a 2-torus (with $r$-dependent conformal factor and
complex modulus) which degenerates into a circle at $r=0$.  The
space  is  actually regular everywhere, including  $r=0$, as 
becomes clear in rewriting the metric in cartesian coordinates.}
 If the 11-th direction is different
from  $y$, this $D=11$ background reduces  to  a similar flat 
 one in $D=10$
 which describes  the NS-NS
Kaluza-Klein  Melvin 
solution upon further reduction to $D=9$
(as in \melvin\ in what follows we shall refer to 
 this flat $D=10$ background  
as 
Kaluza-Klein  Melvin  model).
 If instead we  reduce along $y=x_{11}$ we 
  end  up with   a  curved type IIA  $D=10$ 
background
with a non-trivial RR vector potential representing 
a magnetic flux `tube' (7-brane).

The key technical 
 point is that the   GS string action in this
$D=10$  RR background
  can be   found  by the  double
dimensional reduction \duf\ from the  
explicitly known  supermembrane action \berg\
in the corresponding  {\it flat}  $D=11$ 
space (the relevant  membrane action
is related to the standard 
action  written in  terms of flat
cartesian coordinates by a simple coordinate transformation).\foot{The  
 general  superspace 
  expression for a  GS  superstring  action in 
  an arbitrary type IIA supergravity background was 
  originally derived \duf\ by starting 
  from the membrane action \berg\ 
  in a curved $D=11$ supergravity background.} 
  The resulting  covariant string action is 
  non-trivial, containing higher powers 
  of $\t$'s. After fixing  the light-cone gauge,
  the action  simplifies
  to a form containing only quadratic and 
  quartic terms in fermions.

\bigskip

2. The ten-dimensional background that we shall  consider
 is a 
solution type IIA supergravity  with the relevant part of the 
 action being 
\eqn\actio{
S=\int d^{10}x \sqrt{-G} \bigg( e^{-2\phi}\big[R+4 \big(\del\phi\big)^2\big]
-\ha F_{mn }F^{mn} \bigg) \ .
}
It is given by  ($q=$const)
\eqn\mccc{
ds^2_{10}=f(r)\big[ -dt^2+dx_1^2+...+dx_7^2 +dr^2 +{r^2 f^{-2}(r) } d\vp ^2 \big]\ ,
}
\eqn\mzzz{
\ e^{{2}\phi }=f^{3} (r)\ ,\ \ \ \ A ={qr^2  f^{-2}(r)} d\vp  \ , 
\ \ \ \ \ f \equiv (1+ q^2
r^2)^{1/2} \ ,
}
where $A$ is the RR 1-form.
It can be obtained 
by dimensional reduction  from the  $D=11$ background  with trivial 3-form field $A_{\mu\nu\rho}=0$
and the  metric
\eqn\maaa{
ds^2_{11}=-dt^2+dx_1^2+...+dx_7^2+dr^2+ r^2 (d\vp +  q dy)^2+dy^2 \ . 
}
Here $\varphi \equiv \vp +2\pi $ and $y\equiv x_{11}$ 
has period $2\pi  R_{11}$ so that this  metric is topologically non-trivial if 
$qR_{11}\not=n$.
Since 
the metric is locally flat this  an exact solution to the
equations of motion of eleven-dimensional supergravity {\it and} M-theory
(all possible higher-order curvature corrections  vanish).
If the dimensional reduction is
done along one of the  coordinates $x_1,...,x_7$ 
  one obtains  the $D=10$  metric representing the 
  NS-NS Kaluza-Klein 
Melvin background  discussed in \refs{\gib \gibma \dowo -\gaun  \melvin}.
 Here instead 
we  shall consider the  dimensional reduction   along 
$y$. 
By writing the metric   \maaa\ as
\eqn\mbbb{
ds^2_{11}=-dt^2+dx_1^2+...+dx_7^2+dr^2+ {r^2 f^{-2}(r) } d\vp ^2 
+f^{2}(r)[dy+q r^2 f^{-2}(r) d\vp ]^2 \  , 
}
we obtain the  above type IIA solution \mccc\ with  the magnetic 
RR vector  field.  This RR  background is obviously related
to the NS-NS Melvin solution by U-duality.\foot{Explicitly,
the  Kaluza-Klein Melvin model in type IIA theory is 
obtained from  the RR Melvin model by applying T-duality along $x_7$, then S-duality
in IIB theory, and then T-duality along $x_7$ to get back to IIA theory.
Though the IIA KK Melvin model  described by  flat $D=10$ space  is 
 an exact solution of string theory, this may not apply 
to the RR Melvin model  since due to lack of supersymmetry 
there may be both perturbative and non-perturbative corrections to the duality
transformations. Let us note also that  KK Melvin solution of type IIB 
theory is related to that of type IIA theory by a trivial T-duality transformation
along one of the  flat `spectator' coordinates.}

The background \mccc\ 
 is a curved space-time corresponding to an axially
  symmetric magnetic RR `flux  tube' (7-brane, see below).
 If  the string coupling $e^{\phi_0+ \phi (r)}$  is chosen to be
  small  at  $qr\ll 1$, it becomes 
  large at $qr\gg 1$, so that the geometry is ten-dimensional  
  close to the core of 7-brane (inside the flux tube) and becomes 
eleven-dimensional  far  from it.

It is  useful  to make the following  coordinate transformation:
\eqn\mddd{
 f(r){dr\ov r}={d\rho\ov \rho }\ ,\ \ \ \ \ \ 
 \rho=2 q\inv  e^{f(r)-1} [{f(r)-1\ov f(r)+1}]^{1/2}  \ , 
}
putting the solution into the  explicit 
rotationally symmetric 7-brane form
\eqn\meee{
ds^2_{10}=H_1(\rho ) \big( -dt^2+dx_1^2+...+dx_7^2 \big)
+ H_2(\rho ) \big( d\rho^2 +\rho^2  d\vp ^2 \big)\ ,
}
$$
  e^{2\phi }=H^{3}_1(\rho )\ ,\ \ \ \
A={q \rho^2 H_2(\rho )\ov H_1(\rho )}d\varphi  \  , \ \ \ \ \ 
 H_1\equiv f[r(\rho)]\ ,\ \  \  H_2 \equiv {r^2(\rho)\ov \rho^2 f[r(\rho)] }
  \ . 
$$
Like the U-dual KK Melvin  background, this 
 7-brane  solution  breaks all  supersymmetries.\foot{The $D=11$ Killing
  spinor $\varepsilon$ does not satisfy
the periodic boundary condition in $y$ (unless $qR_{11}=2 n$)  since 
$
\varepsilon(x,y+2\pi R_{11})=\exp [{ 1 \ov 2} 
{qR_{11}} \g_8 \g_{9} ]\ \varepsilon(x,y), 
$ where $8,9$ correspond to the directions in the $(r,\vp)$ plane.}
The behavior at small and large $\rho $ is  as follows:
$$
r(\rho )\cong \rho \ (1-\four q^2\rho^2) \ ,\ \ \ 
H_1(\rho)\cong 1+\ha q^2\rho^2\ ,\ \ \ \ H_2(\rho )\cong 1-q^2\rho ^2\ \ \ \ \ \ {\rm if}\ 
q\rho \ll 1
$$
$$
q\ r(\rho )\cong \log (q\rho ) \ ,\ \ \ 
H_1(\rho )\cong \log (q\rho)\ ,\ \ \ \ H_2(\rho )\cong { \log(q\rho)\ov q^2\rho^2 }
\ \ \ \ \ \ {\rm if}\ q\rho \gg 1 \ . 
$$
The gauge field  strength $dA $ is constant near  the origin 
and asymptotically approaches zero  at large $\rho$.

Our aim below will be to find the exact expression for the action of 
the GS string 
propagating in the background \mccc,\mzzz.
We shall obtain it by double dimensional reduction 
 along $x_{11}=y$
from the $D=11$  action  of a  membrane moving in the metric \maaa.

The  reduction  of the same  membrane  action along the `trivial' direction
 $x_7 $ leads to the  action for GS string in the Kaluza-Klein Melvin background.
In the light-cone gauge, the   non-trivial part of the latter action
 (without terms corresponding to free  bosonic directions 
 $x_1,...,x_6$)
   has the form  \melvin\foot{It is useful 
 to recall also that under T-duality along $y$ 
 the KK Melvin  background  is transformed into 
 a curved  background described by the string model 
 with the following bosonic part (this is a special case of a 3-parameter 
 class of magnetic models solved in \rutse):
 $
\td L= \del_+ r \del_- r +  F(r)r^2 (\del_+ \vp  + q \del_+ \td y)
( \del_-\vp
-q \del_-  \td y) + \del_+ \td y \del_- \td y  $ $
+   {\cal R} (\p_0 +  \ha \ln F )$ , \
$  F\equiv (1 + q^2 r^2)\inv,$
    \ ${\cal R}\equiv \four \a' \sqrt{g} R^{(2)} .$
This model
is  equivalent to  the KK Melvin model
 at the CFT level,  i.e.  it has, in particular,
  the same  mass spectrum \melvin.} 
\eqn\kkm{
L_{\rm KK\ Melvin}
=    \del_+ y \del_- y
 + D_+ x_s D_- x_s + i S_R \DD_+ S_R + i S_L \DD_- S_L 
\ , }
\eqn\pop{ {(D_a)}_{st} \equiv \delta_{st} \del_a  - q \ep_{st} \del_a y \ , 
\ \ \ \ \ \ 
\DD_i  \equiv  \del_a  - \four  q \ep_{st} \g_{st}  \del_a y \  , \ \ \ \ \  s,t=8,9 \ .  }

\bigskip


3. The action  for  the $D=11$ supermembrane 
coupled to a $D=11$ supergravity background is given by \berg\
\eqn\mem{ I= 
- \ha T_3  \int d^3 \xi \bigg[\sqrt {-g} ( g^{ i j}
 \h \Pi^{\hat \m} _i \h \Pi^{\hat \n}_j \eta_{\hat \m \hat \n} -1)
- \third 
 \ep^{ijk} \h \Pi^A_i\h \Pi^B_j \h \Pi^C_k A_{CBA} \bigg] \ . }
Here $g_{ij}$ is the auxiliary 3d metric 
($i=1,2,3$), \ 
$\h \Pi^A_i = \del_i Z^M \h E_M^A (Z)$, \  
$Z^M=(x^\m, \t^\a)$ \ ($\m= 0,1,...,9, 11$, $\a=1,..., 32$), 
\ $ A_{CBA}= A_{CBA} (Z)$
($A= (\hat \m, \hat \alpha)$ are superspace tangent indices).\foot{We shall  assume that $\g^\m$ 
($\g_\m\g_\n + \g_\n\g_\m = $diag$(-1,1...,1)$)
are  chosen to be  real   and  the  Majorana  spinors  $\t$ are real, 
$\bt = \t^T  C, \ C=\g^0$. Note that 
$C_{\a\b}, (C\g^{\m_1\m_2\m_3})_{\a\b}, 
(C\g^{\m_1\m_2\m_3\m_4})_{\a\b}$
are antisymmetric while
 $(C\g^\m)_{\a\b}, (C\g^{\m_1\m_2})_{\a\b}, 
(C\g^{\m_1\m_2\m_3\m_4\m_5})_{\a\b}$
are symmetric in $\a,\b$.
In particular, 
$\bt \g^\m\t=0$.}
As  was shown in \refs{\berg,\duf},  this  
 action is invariant under $\kappa$-symmetry 
 provided  the background  satisfies  the 
 superspace equations of on-shell $D=11$ supergravity
  \refs{ \crem,\bri}. 

In flat superspace  the action \mem\ takes the following explicit form
\berg\
\eqn\memf{
 I= - \ha T_3  \int d^3 \xi \bigg[\sqrt {-g} g^{ i j} 
 (\del_i x^\m - {i} \bar\t \g^\m\del_i \t)
 (\del_j x^\m - {i} \bar\t \g^\m\del_j \t) -  \sqrt {-g}
 }
 $$
 +\  i \ep^{ijk}
 \bar \t \g_{\m\n} \del_i \t (\del_j x^\m  \del_k  x^\n 
  -  i \del_j x^\m \bar\t \g^\n\del_k \t
  - \third \bar\t \g^\m\del_j \t \bar\t \g^\n\del_k \t )
  \bigg] \equiv I_N + I_{WZ} \  . 
  $$
 Eliminating the auxiliary metric,  
 the  first `even' term of this action can be written in the
 standard form 
 \eqn\dira{
 I_N = - T_3  \int d^3 \xi  \sqrt {- {\det}_3 h_{ij} } \ , 
 \ \ \ \ \ \ 
 h_{ij} =  (\del_i x^\m - {i} \bar\t \g^\m\del_i \t)
 (\del_j x^\m - {i} \bar\t \g^\m\del_j \t) \ . }
 Let us now  consider the  background \maaa.
 The   metric \maaa\ is flat 
 since locally it  can be  
obtained from the standard cartesian 
 metric of  $R^{1,7} \times R^2 \times S^1$ 
by a  $y$-dependent  rotation
in the  $(x_8,x_{9})$  plane.
Therefore, 
the corresponding  supermembrane action is  formally
related to the  flat space membrane action \memf\ 
 by  the  substitution 
\eqn\raaa{\t  \to 
\t_\ff = e^{- { 1 \ov 2}  q \g_{*} y}\ \t\ ,\ \ \ \ \ \ \
\ \ \g_{*}\equiv \g_8 \g_{9}\ ,
}
$$x^8 +  ix^{9} = r e^{i\vp}\   \to \ 
x^8_\ff  +    ix^{9}_\ff = e^{ iqy} (x^8  +   ix^{9})\ ,\ \ \ \ 
\ x^\mu_\ff=x^\mu\ ,\ \ \mu\neq  8,9 \ ,  
$$
 where $(x^\mu, \t^\a )$ are the `true' 
  superspace coordinates with  correct periodicity 
(the `flat'  coordinates  are 
 not  single-valued in the $y$ direction).
 The action \memf\ 
 then becomes 
 \eqn\memfi{
 I=  - T_3  \int d^3 \xi  \sqrt {- {\det}_3 h_{ij} } +     I_{WZ} \ , }
 \eqn\hhh{ 
 h_{ij}=
 (D_i x^\m - {i} \bar\t \g^\m\DD_i \t)
 (D_j x^\m - {i} \bar\t \g^\m\DD_j \t)  \ , }  
   \eqn\wwz{
 I_{WZ}=
  - \ha i T_3 \int d^3 \xi  \   
  \ep^{ijk}
 \bar \t \g_{\m\n}  \DD_i \t 
 (D_j x^\m  D_k  x^\n
  -  i  D_j x^\m \bar\t \g^\n\DD_k \t  
  - \third \bar\t \g^\m\DD_j \t \bar\t \g^\n\DD_k \t )  \ ,
  }
 where
 the covariant derivatives  are defined as follows (they are  the same as 
 in \pop)
 \eqn\deri{ 
 D_i x_s \equiv  \del_i x_s  - q \ep_{st} x_{t }\del_i y \ , 
 \ \ \ \   D_i x_u \equiv  \del_i x_u \ , \ \ \ \ 
 D_i x_{11}= D_i y \equiv  \del_i y \ , \ \ \ 
  } $$ 
\DD_i \t  \equiv  (\del_i  - \ha  q  \g_*  \del_i y)\t  \  ,  $$
where 
 $s,t=8,9$   and   $u= 0,1,..,7$.

In order to obtain the type IIA GS superstring  action 
in  $N=2, D=10$ supergravity background 
\mccc,\mzzz\ we perform the 
double dimensional reduction \duf\ by splitting the  
world-volume and space-time coordinates as   
$\xi^i\to (\xi^{a},\xi^3)$, 
 \ $x^\m\to (x^{{m}},x^{11}\equiv y)$, 
 assuming
that $\del_3 g_{ij}=0$, $\del_3 x^{m} =0,$ \  $\del_3 \t=0$, 
and
setting 
$y=R_{11}\xi^3$. 
Here and in what follows
$a,b=1,2$, \ $m,n= 0, 1,..., 9$
(repeated indices are summed with Minkowski metric).

The WZ term \wwz\ then reduces to  ($T_2=2\pi R_{11}T_3 $)
%
%
\eqn\wzwz{
I_{WZ}=I^{(0)}_{WZ}+ I^{(q)}_{WZ} \ ,
}
\eqn\wzzz{
I^{(0)}_{WZ}=-\ha i T_2\int d^2\xi \  \ep^{ab} \bt \g_{m}
\G_{11} \del_a \t (  2 \del_b x^m  - i \bt \g^m \del_b \t)  \ ,
}
$$
I^{(q)}_{WZ}=-\ha i q T_2\int d^2\xi \  \ep^{ab}\  \bigg[\ \ep_{ts} x_s
\bt \G_{t m} \del_a\t\  (2 \del_b x^m-i \bt \G^m \del_b\t ) -i
\ep_{ts} x_s \bt \G_{t} \g_{11} \del_a\t\  \bt \g_{11} \del_b\t 
$$
$$
-\ \ha \bt \G_{mn} \G_*\t (\del_a x^m \del_b x^n - i\del_a x^m \bt \G^n
\del_b\t - {\textstyle {1\ov 3} }\bt \G^m\del_a\t \bt \G^n \del_b\t )
$$ 
$$
 - \ \ha \bt \G^n  \G_*\t \ \bt \G_{mn} \del_b \t  (  i\del_a x^m  -
{\textstyle {2\ov 3} }\bt \G^m \del_a\t  ) 
$$
$$
+\  \ha \bt \G_{m}\G_{11} \G_*\t (i\del_a x^m \bt \G^{11} \del_b\t +
{\textstyle {2\ov 3} }
\bt \G^m\del_a\t \ \bt \G^{11} \del_b\t )
$$
\eqn\wwzz{
 - \ \ha \bt \G_{m}\G_{11} \del_b \t  ( i\del_a x^m  \bt \G^{11}\G_*\t \  
- {\textstyle {2\ov 3} }\bt \G^{11}  \G_*\t \ \bt \G^m \del_a\t  
+{\textstyle {2\ov 3} }\bt \G^{m}  \G_*\t \ \bt \G^{11} \del_a\t  )
\bigg]\ .
}

Let us now consider the more complicated 
induced metric  term $I_N$ (the first term  in \memfi). 
In general,
\eqn\grt{
{\det}_3 h_{ij}  = {\det}_2 f_{ab}   \ , 
\ \ \ \ \ \ \
f_{ab} = h_{33}^{1/2}( h_{ab}  -  h_{33}^{-1}  h_{a3} h_{b3} ) \  ,  
}
where 
\eqn\comp{
h_{ab} 
= (\del_{a} x^{m} - {i} \bar\t \g^{m}\del_{a} \t)
 (\del_{b} x^{m}- {i} \bar\t \g^{m} \del_{b} \t)
  - \bar\t \g_{11}\del_{a} \t
  \bar\t \g_{11} \del_{b} \t \ , }
 $$
 h_{a 3}  =  R_{11} \bigg(- i \bar\t \g_{11}\del_{a} \t  +  
  q \big[ \ep_{ts} x^t(\del_{a} x^s - {i} \bar\t \g^s\del_{a} \t)
    $$ $$ 
 +\ 
\ha  {i} (\del_a x^u - {i} \bar\t \g^u\del_a \t)
   \bar\t \g^u\g_* \t  
     +  \ha   \bar\t \g_{11}\del_a \t
   \bar\t \g_{11} \g_* \t\ \big]\bigg)  \equiv R_{11} \bar h_{a 3}\ , 
$$
$$
h_{33} = 
(R_{11})^2 \bigg[ (1 +\ha  {i}  q \bar\t \g_{11}\g_*  \t)^2
 +  q^2  x_s x_s 
 - \four q^2  ( \bar\t \g^u \g_* \t )^2     
      \bigg] \equiv (R_{11})^2 \bar h_{33} \ ,  
 $$
 where we have used that $\bt \g^\mu \t =0$.
The 
 corresponding   part of the string 
 action   $\int d^2\xi \sqrt {- \det_2 f_{ab} }$
 can be represented  in the usual 
 form  with an auxiliary 2d metric $g_{ab}$:
 \  
$\ha  \int d^2\xi \sqrt {-g} g^{ab} f_{ab}$.

 Choosing  
  the standard 
conformal gauge $ \sqrt {-g} g^{ab} = \eta^{ab}$
we  finish with the following GS action:
$I=I_N + I_{WZ}$, where  $I_{WZ}$ 
was given in \wzzz,\wwzz,  and 
\eqn\even{
I_N =  - \ha T_2 \int d^2\xi\ 
 \bar h_{33}^{1/2}\ ( h_{aa}  - 
  \bar h_{33}^{-1}  \bar h_{a3} \bar h_{a3} ) \ , }
 where the repeated indices are contracted with the 
 flat 2d metric.
 It is easy to check that the bosonic terms in this action 
 have  the  standard $\sigma$-model form  $G_{mn} (x) \del_a x^m \del_a x^n$
   corresponding to the metric \mccc. 
  Indeed, note that
 $x_8= r \cos \vp, \ x_{9} = r \sin \vp,$  and 
 $$\bar h_{33}  =   f^2(r) + O(\t^2) \ , \ \ \ \ \ \   dx_s dx_s 
   -  f^{-2} (r)  (\ep_{st}  x_s  d x_{t})^2
 = dr^2 +  f^{-2} (r)  r^2 d \vp^2 \  . $$

 In the  $q=0$ limit  the action  reduces to  the standard  expression for the GS
 action in the  flat $D=10$ space   \gs
 $$I(q=0) =  - \ha T_2 \int d^2\xi\ \bigg[
 (\del_{a} x^{m} - {i} \bar\t \g^{m}\del_{a}
 \t)
 (\del_{a} x^{m}- {i} \bar\t \g^{m} \del_{a} \t) \  $$
 \eqn\gree{
 + \ i \ep^{ab} \bt \g_{m} \G_{11} \del_a \t (  2 \del_b x^m 
 - i \bt \g^m \del_b \t) 
 \bigg]\ ,  }
 written here in the   $D=11$ notation for the   Majorana  spinors.\foot{
 Relation to   the two $D=10$  Majorana-Weyl  spinors
is  $\t= (\t^1,\t^2), \ \G_{11} =$ diag$ (I,-I)$, \
 $\Gamma^m = \pmatrix{ 0 & \gamma^m \cr  \gamma^m & 0}$.}
 For $q\not=0$  the action has a 
 complicated    form, 
especially because 
of  the $\t$-dependent 
$h_{33}^{1/2}$ factor in \even\ 
(which is equal to 
$e^{{2\ov 3}\Phi(x,\t)}$  where $\Phi$  
  is  the $D=10$ dilaton superfield).\foot{The explicit expressions for
  the $D=10$ superfields representing  the supergeometry of our
  background can be read off from the  superstring action
  by comparing it to the general result \duf\
  of the  double dimensional reduction of the membrane action in curved
  $D=11$ space.}
 The presence of  higher-order  fermionic terms 
 in the action 
reflects the curvature of the background. 
 The WZ term contains 
  terms which are of  first order only in 
 the magnetic flux parameter $q$
 which   describe, in particular, 
   the coupling to the  RR   vector field strength.\foot{
   In particular,  \wwzz\  contains  the term 
   $ e^\p F_{st} \ep_{ab} \del_a x^m \del_b x^n   \bt \g_{mn} \g_{st} \t $\ 
   (where $F_{st}$ is the RR vector field strength) 
   expected  on the general grounds (e.g., from 
   comparison with the RR vertex operator).
   For $m,n=8,9$ this is the term derived from $D=11$
   membrane action 
   in   \tsey.} 
 Since the action  explicitly depends on $\t$
 (and not only on $\del_a \t$) 
 it does not have  obvious global translational fermionic   symmetries
 (indeed, 
the magnetic background breaks all supersymmetries).
 
4. \  
The action simplifies dramatically 
 once one  fixes  the  light-cone gauge.
Since the action is $\k$-symmetric  we are free 
to impose the condition 
\eqn\lco{\g^+ \t=0\ , \ \ \ \ \   \ \ \ \ \   
\g^\pm  = { 1 \ov \sqrt 2} (\g^0 \pm  \g^1)  \ .  }
Note that our  background is non-trivial only in the $8,9$ 
directions, i.e.  the choice of the light-cone gauge `commutes' with the rotation 
\raaa\ (in particular, $[\g^+, \g_{\m}]=0$ for $\m= 8,9,11$). 
Instead of using \lco\ directly  in the string action which follows
from  \memfi\
we may  obtain the same  final expression  in  a  
more straightforward  way 
by first   choosing the  fermionic light-cone gauge 
 in the free membrane action \memf\
(assuming $\del_3 \t=0$) 
and {\it then} applying  the rotation \raaa\ and  dimensional reduction.
The standard observation is that if $[{\cal O}, \g^+] =0$ and $\t$
satisfies  \lco\ then 
 $\bt \g^\m  {\cal O}  \t$ 
is equal to zero unless the index $\m$ takes the 
value $-$.\foot{As usual, 
for Majorana spinors 
$\bar \t = \t^T C ,$ \ $ 
C \g_m = - \g^T_m C,$  so that 
$\g^+ \t=0 $ implies $    \bar \t \g^+ =0$.
One is also to note that 
 $1 = \g^+ \g^- + \g^- \g^+$ may be inserted in the fermionic bilinears.}
This implies, in particular, that 
all  quartic fermionic terms in $h_{ab},h_{a3}$ 
and all fermionic terms in $h_{33}$ in \comp\
vanish,  so that $\bar h_{33} $ becomes  $\t$-independent and equal to  
$f^2
= 1 + q^2 x_s x_s$. 
Using \lco \ in the  flat membrane action \memf\ 
and then making the rotation (i.e. replacing the derivatives by the covariant
derivatives \deri)
we find the action  which contains only terms which are at most 
 bilinear in fermions, 
$$
 I= - \ha T_3  \int d^3 \xi \bigg[\sqrt {-g} g^{ i j} 
 ( D_i x^\m  D_j x^\m -  2 {i} \bar\t \g^- \DD_i \t
 \del_j x^+)  -  \sqrt {-g}
  $$\eqn\memfl{
 +\  2i \ep^{ijk}
 \bar \t \g^{p} \g^{-} \DD_i \t  D_j x^p  \del_k  x^+ 
  \bigg] \  , 
  }
  where $p=2,...,9, 11$. 
  The term  with $p=11$ in the  WZ  part  of \memfl\   leads  to 
  the standard  WZ  term 
  $
2 i\ep^{ab}\bt  \g^- \g_{11} \del_a \t \del_b x^+ $
in the light-cone gauge  GS string action.

 The elimination of  the auxiliary metric  after  dimensional reduction 
 reintroduces a  quartic fermionic term. Equivalently, 
  a $\t^2$ term in $h_{a3}$ in \comp\
  gives rise to the $\t^4$ term in \even. As follows from \comp,\lco\ 
 $$
h_{ab} = \del_a x^m \del_b x^m  - 2 i \del_a x^+ \bar \t \g^- \del_a \t \ ,
\ \ \ 
\bar h_{a3} = q  (\ep_{ts}  x_t \del_a x _s  + \ha i \del_a x^+ \bar\t  \g^- \g_*
\t  )\ , $$ 
and $\bar h_{33} =1 + q^2 x_sx_s = f^2 (r).$
 The light-cone  string  model  we get  is thus  the following
$$
I=-\ha T_2 \int d^2\xi \bigg[
f(r) ( \del_a x^m \del_a x^m - 2 i\del_a x^+ \bt \g^- \del_a \t   ) 
$$ $$
- \ q^2 f^{-1} (r)  ( \ep_{ts} x_t \del_a x_s  + \ha i\del_a x^+  \bt \g^-  
\g_* \t )^2 
$$ \eqn\lcac{ + \ 2 i\ep^{ab}\del_b x^+ \big[ 
\bt \g^-  \g_{11} \del_a \t  
-      q  (\ep_{st} x_t \bt  \g^- \g_s\del_a \t    
   + \ha \del_a x^p \bt  \g^- \g_* \g^p \t    ) \big] 
    \bigg]
   \ .}
The $\t^4$ term reflects the curvature of the background, while the 
 RR coupling  term  is only quadratic in fermions.\foot{In the case of   the 
KK Melvin (NS-NS) model 
 similar procedure of starting  with a covariant  superstring (or
 membrane)   action, 
 fixing light-cone gauge and  then performing the 
 rotation in the plane 
 gives the action \kkm\ 
  with the fermionic part 
$
-2i \del_a x^+  \bar \t \g^-  ( \delta_{ab} - \g_{11} \ep_{ab} ) 
\DD_b \t\  $.
It is easy to check  explicitly that the procedures of rotation and 
light-cone gauge fixing indeed commute.}

Next, one may  fix the  remaining 
conformal invariance  by the bosonic light-cone gauge 
$x^+= c  + p^+ \xi^1$ (the 7-brane  background 
\mccc,\mzzz) 
 has $SO(1,7)$ Lorentz
symmetry). This gauge choice  could  actually
 be  made  already  at the level of the membrane action \memfl\  (as in the standard flat space 
case \berg) since it `commutes' with the rotation in the $(8,9)$ plane.
In contrast to the  flat   KK Melvin model  
case \kkm, however, 
the action \lcac\ 
does not   take a  manifestly 
2d Lorentz-covariant form with $\t$ 
replaced by a pair of 2d spinors 
(the 2d parity even and odd terms  in \lcac\
do not combine as they do in the case of a flat 
target space since 
the even term here 
  has  the  extra factor of $f(r)$).
Nevertheless, since the  action is 
 at most quartic in fermions, 
it may be  useful   to perform 
 a triality transformation 
on spinors  to    convert them 
into $SO(8)$ vectors,
and that  may   lead to a more 
 convenient framework 
 to demonstrate the conformal invariance 
of this model  at the quantum level.

The flat space 
 membrane action  describes   a non-trivial   interacting 
model even after fixing   the light-cone  gauge \refs{\berg,\dewi}.
While the double dimensional reduction of it    along one of the `spectator' 
$x^m, \ m=1,...,7$ coordinates 
leads to a simple quadratic action \kkm\   of KK Melvin  
model (which can be  solved in terms of free oscillators
 by `undoing' the rotation in the plane \melvin),
 the reduction along $y$ leads 
to the non-trivial interacting  theory \lcac , which can no longer be put 
into a flat or quadratic 
form by any obvious redefinition of the supercoordinates.

5. \ Similar considerations based on double dimensional 
reduction of the  $D=11$ membrane action 
 allow one to find the form of the  GS 
superstring action in other
type IIA backgrounds which can be obtained by 
reduction from  locally {\it  flat }
$D=11$ backgrounds. 
One particularly  interesting  example is the RR background  representing 
the near-core 
region 
of a D6-brane. It  can be obtained  by reduction from  the 
near-core part of the  KK monopole 
\towns\ which happens to be a flat space --
an ALE space
 with  $A_{N-1}$ singularity  times a  7-dimensional 
Minkowski space $M^{(6,1)}$
($S^3$ part of $R^4$ is  replaced by a Hopf
  fibration $S^2\times S^1$). 
The Green-Schwarz  action for a string in the   
 curved  magnetic RR background  corresponding to 
 the near core (or decoupling \mald)  region of the D6-brane can be thus obtained by
 the same method as used above. 
 Starting with the membrane action 
 in the locally flat $D=11$  background
$$ ds^2_{11} =- dt^2 + dx^2_1 + ...+ dx^2_6+ 
d\rho^2 + \rho^2 (d\tilde \p^2 +
  {\rm sin}^2\tilde \p\ d\tilde\varphi^2 +
{\rm  cos}^2\tilde \p\ d\tilde \psi^2 )
  $$
 $$
  = - dt^2 + dx^2_1 + ...+ dx^2_6
 +   {\ha Nr\inv dr^2}+ \ha N  r 
 (d\p^2 + {\rm sin}^2 \p\ d \vp^2 )$$
$$  +\    {2  N\inv}r
[d\psi  + \ha N (\cos \p -1) d\vp]^2\  ,  
 $$
 where $ r\equiv \ha N\inv  \rho^2$, 
 $\p= 2\td \p, \ \vp= \td \vp - \td \psi , \ \psi = N \td \psi \equiv \psi +
 2 \pi,$
 and making the double dimensional 
 reduction  along $x_{11}= R_{11} \psi$ 
  leads to the covariant Green-Schwarz action
  for this background.
  We expect that  the resulting action will be
 simpler than    \lcac ,  
 since the basic  function of  the radial coordinate 
   here  is  just a power of $r$.

  \bigskip\bigskip
  
\bigskip

We are  grateful  to  N. Berkovits,  B. de Wit,   R. Kallosh and 
  R. Metsaev
for useful discussions  and  comments on the paper.
 We also  acknowledge  the support
 of PPARC and  the European
Commission TMR programme grants ERBFMRX-CT96-0045 and
ERBFMBI-CT96-0982.

\vfill\eject
\listrefs
\vfill\eject\end